\documentclass{osa-article}
\journal{osajournal}

\articletype{Research Article}

\begin{document}

\title{Spherical sampling methods for the calculation of metamer mismatch volumes.}

\author{Michal Mackiewicz,\authormark{1,*} Hans Jakob Rivertz,\authormark{2} and Graham Finlayson\authormark{1}}

\address{\authormark{1}School of Computing Sciences, University of East Anglia, Norwich, UK\\
\authormark{2}Department of Computer Science, Norwegian University of Science and Technology, Trondheim, Norway\\
}

\email{\authormark{*}M.Mackiewicz@uea.ac.uk} 



\begin{abstract}
In this article, we propose two methods of calculating a theoretically maximal metamer mismatch volumes. Unlike prior art techniques, our methods do not make any assumptions on the shape of spectra on the boundary of the mismatch volumes.  Both methods utilise a spherical sampling approach, but they calculate mismatch volumes in two different ways. The first method uses a linear programming optimisation, while the second is a computational geometry approach based on half-space intersection. We show that under certain conditions the theoretically maximal metamer mismatch volume is significantly larger than the one approximated using prior art method.
\end{abstract}

One print or electronic copy may be made for personal use only. Systematic reproduction and distribution, duplication of any material in this paper for a fee or for commercial purposes, or modifications of the content of this paper are prohibited.

\homepage{https://doi.org/10.1364/JOSAA.36.000096}

\graphicspath{{.}{./figs/}}
\section{Introduction}

A spectral power distribution carries the complete information about the light. However, the human visual system (HVS) or a typical camera has only three colour receptors or sensors, which means that colour information is compressed to only three quantities. Human eye as well as typical camera sensors collect light over broad range of wavelengths and this implies that different spectral power distributions can be mapped to the same set of three colour responses. Consequently, there will be physical samples which will reflect lights of different spectral power distribution that will produce the same colour response under one light and different colour responses under another light \cite{WYSZECKI}. Analogously, if we keep illumination unchanged, but change the observer (human or camera), the same phenomenon may occur i.e. two objects can produce the same colour response for one observer and different colour responses for another. This phenomenon is called metamer mismatching and in this paper we will discuss the issue of the extent of such colour mismatch occurrences. More precisely, given the observed colour match under the first condition, we consider the range of possible colour responses that the two objects can produce under the second condition. The set of all such colours is called the metamer mismatch volume.

Metamer mismatching is important from the viewpoint of camera sensor design. Colour sensors producing large mismatch volumes for the change of observer from camera to a human are highly undesirable. Nevertheless, modern cameras exhibit some level of metamer mismatching necessitating colour correction algorithms in the camera processing pipeline. Development of such algorithms can be aided by a better understanding of the phenomenon in question. Lighting design is another field  where metamer mismatching plays an important role, particularly in the context of rapidly growing modern LED lighting. Finally, understanding mismatch volumes helps us understand our own vision and therefore it is an interesting question in its own right that was studied from the the emergence of the colour science.

Recent approaches to calculation of the metamer mismatch volumes were underpinned by the assumption that reflectance spectra of all objects can be modelled using a low dimensional linear model \cite{FINLAYSON.MOROVIC.99, FINLAYSON.MOROVIC.05, URBAN04}. The reliance on such a model results in only approximate estimation of the real mismatch volume, notably such estimates will always underestimate the size of these objects. More recently, Logvinenko et al. in \cite{LOGVINENKO.FUNT.TIP14} proposed to address this problem by calculating the full extent of the metamer mismatch volumes by means of establishing their precise boundaries. This approach is motivated by the observation that there is no metamerism at the boundary of the mismatch volume i.e. a colour response at the boundary of the mismatch volume corresponds to a unique spectral reflectance function. Logvinenko et al. began by asking a question: what is the general form of the shape of the reflectance spectra on the boundary of the mismatch volume? They noted that these are elementary step functions of zeros and ones with some limited number of transitions between these two values. They parametrised these functions with respect to the wavelengths where the transitions between 0 and 1 occur and decided to model them with up to five transitions. They \textit{conjectured} that such a model should provide a good approximation of the theoretically maximal mismatch volumes. However, as we will later show in our experiments, their chosen parametrisation results in significant underestimation of the theoretically maximal volumes of these objects.

The rest of the paper is organised as follows. In Section \ref{sec:theory}, we introduce the relevant theory of mismatch volumes and notation and briefly describe the Logvinenko et al.'s algorithm \cite{LOGVINENKO.FUNT.TIP14}. In Section \ref{sec:calc_OCS}, we describe our early related work \cite{MACKIEWICZ16.PICS} that underpins our methods, which are subsequently described in Sections \ref{sec:calc_lin_prog} and \ref{sec:wout_optimisation}. The experiments evaluating the effectiveness of our algorithms and the comparison with the prior art can be found in Section \ref{sec:exp_res}. 

The preliminary version of our first algorithm was presented at a conference \cite{MACKIEWICZ17.CIC}. Here, we present an extended version of this work, which includes the second and significantly more efficient algorithm which does not require optimisation. We also give additional results and discussions.

\section{Metamer Mismatch Volumes Theory}\label{sec:theory}

The recent work by Logvinenko et al. in \cite{LOGVINENKO.FUNT.TIP14} includes an extensive introduction to metamer mismatching theory. We summarise their major points below.

The colour responses of the set of $N$ sensors $\Phi(r)=(\phi_1(r),\phi_2(r),...,\phi_N(r))$ to an object with a spectral reflectance function $r(\lambda)$ illuminated by the light with the spectral power distribution $e(\lambda)$ are given by the colour formation equation:

\begin{equation}
\phi_i(r)=\int_{\lambda_{min}}^{\lambda_{max}}r(\lambda)e(\lambda)c_i(\lambda)\,\mathrm{d}\lambda,\quad i=1,2,..,N.
\label{eq:col_for}
\end{equation}

where $\lambda_{min}$ and $\lambda_{max}$ denote the limits of the visible spectrum and $c_i$ are the sensor spectral sensitivities. A reflectance spectrum $r(\lambda)$ is a function with values between zero and one. 

For the human visual system, $c_i(\lambda)$ would denote cone fundamentals or colour matching functions \cite{WYSZECKI} and hence $N$ is 3. Most modern digital cameras would also use three colour sensors in an attempt to emulate the colorimetry of the tristimulus HVS.

The two objects with reflectance functions $r(\lambda)$ and $r'(\lambda)$ are called metameric if they produce the same colour signals $\Phi(r)=\Phi(r')$. Equation \ref{eq:col_for} tells us that if either illumination or the sensor spectra change these two objects may no longer remain metamers. We call the above two cases \textit{illumination-induced} or \textit{observer-induced metamer mismatching} respectively. From Eq. \ref{eq:col_for} it is clear that metamer mismatching is determined by the product of the illuminant spectrum and the sensor sensitivities, which we will call the \textit{colour system} i.e. $\mathbf{s}(\lambda) = \mathbf{c}(\lambda)e(\lambda)$. We substitute the above into Eq. \ref{eq:col_for} resulting in:

\begin{equation}
\Phi(r)=\int_{\lambda_{min}}^{\lambda_{max}}r(\lambda)\mathbf{s}(\lambda)\,\mathrm{d}\lambda.
\label{eq:col_for2}
\end{equation}

Let us consider another set of colour responses $\Psi=(\psi_1,..,\psi_N)$ corresponding to the colour system spectra $\mathbf{s}'(\lambda) = (s_1'(\lambda),...,s_N'(\lambda))$. This new colour system might have resulted from the alteration of the illuminant spectrum, sensor sensitivities or in general both. Logvinenko et al. \cite{LOGVINENKO.FUNT.TIP14} point out that both $\Phi$ and $\Psi$ can be considered as linear maps: $\mathcal{X}\rightarrow \mathbb{R}^N$, where $\mathcal{X}$ denotes the set of all reflectance functions. The sets of all possible colour system responses of either $\Phi$ and $\Psi$ form closed convex sets in $\mathbb{R}^N$. These sets are referred to as \textit{object colour solids} (OCS).

A \textit{metamer set} with respect to the colour system $\Phi$ is defined as the set of all reflectances metameric to a given reflectance $r_0$ i.e. $\Phi^{-1}(\Phi(r_0))= \{r\in \mathcal{X}|\Phi(r)=\Phi(r_0)\}$. In general, the above metamer set will be mapped by $\Psi$ to a non-singleton set which is referred to as a \textit{metamer mismatch volume}.

Logvinenko et al. \cite{LOGVINENKO.FUNT.TIP14} introduced an additional linear map $\Gamma:\mathcal{X}\rightarrow \mathbb{R}^{2N}$ such that $\Gamma(r)=(\mathbf{z},\mathbf{z}')$, where $\mathbf{z}=\Phi(r)$ and $\mathbf{z}'=\Psi(r)$. Consequently, $\Gamma(\mathcal{X})$ is an object colour solid in $\mathbb{R}^{2N}$. The authors observed that for the colour response $\Phi(r_0)=\mathbf{z}_0$, the metamer mismatch volume $\mathcal{M}(\mathbf{z}_0,\Phi,\Psi)=\Psi(\Phi^{-1}(\mathbf{z}_0))=\{
\mathbf{z}' \in \mathbb{R}^N | (\mathbf{z}_0,\mathbf{z}')\in\Gamma(\mathcal{X})\}$ is a cross-section of $\Gamma(\mathcal{X})$.

The points in the interior of the OCS $\Phi(\mathcal{X})$ represent different \textit{metameric classes} where each class maps to a metamer set comprising infinitely many spectra. The points on the boundary of the OCS are different. Here, each point has only one corresponding reflectance spectrum which we will call \textit{optimal}. The main property of the optimal spectra is that they are elementary step functions of zeros and ones.

Schr{\"o}dinger proposed that the optimal spectra on the boundary of the OCS have no more than two  transitions (we call these $m<3$ spectra) \cite{SCHRODINGER20}. West \& Brill later showed that, the ($m<3$) spectra are optimal if and only if the spectrum locus of the chromaticity diagram is strictly convex and well-ordered in wavelength \cite{WEST.BRILL83}. Another mathematical description of the optimal spectra that would also generalise to any object colour solid was delivered by Logvinenko in \cite{LOGVINENKO09} who proposed that for the set of $N$ colour systems the optimal spectra are the elementary step functions with the transition wavelengths at $\lambda_1,...,\lambda_m$ if and only if the above set of transition wavelengths are the only zero-crossings of the following equation:

\begin{equation}
k_1s_1(\lambda)+k_2s_2(\lambda)+...+k_Ns_N(\lambda)=0,
\label{eq:ks}
\end{equation}

where $k_1,k_2,...,k_N$ are a set of arbitrary real numbers, where at least one of them is not equal to zero. 

In \cite{LOGVINENKO.FUNT.TIP14}, Logvinenko et al. extended this description from object colour solids to metamer mismatch volumes. The metamer mismatch volume is denoted as $\mathcal{M}(\mathbf{z}_0,\Phi,\Psi)$ and its boundary as $\partial\mathcal{M}$. The spectra on the boundary of the mismatch volume are called $\mu\mathit{-optimal}$. The authors noted that Eq. \ref{eq:ks} determines the number of transitions in the optimal spectra on the boundary of $\Gamma$ and in the $\mu$-optimal spectra on the boundary of $\mathcal{M}$. Despite that the number of transitions may be large, they chose to approximate the boundaries of the above volumes using optimal or $\mu$-optimal spectra that were constrained to be the elementary step functions with up to five transitions ($m<6$). Their choice was most likely motivated by two reasons. First, their parametrisation of the optimal spectra is based on the finite list of wavelengths where the transitions occur. This required the choice to be made how to limit the length of that list of parameters. Analogously to the 3-D HVS OCS boundary conjectured to be described with the optimal spectra with up to $3-1=2$ transitions, they proposed to approximate the boundary of the larger 6-D $\Gamma(\mathcal{X})$ and its cross-section $\mathcal{M}$ with the optimal spectra with up to $6-1=5$ transitions. Following the aforementioned authors, we will denote the above two boundary approximations as $\partial\Gamma(\mathcal{O}_5)$ and $\partial\mathcal{M}_5$ respectively. 

The idea of describing mismatch volume boundary with $m<6$ elementary step functions has already been discussed in literature. Ohta and Wyszecki held the (incorrect) view that the optimal spectra on the boundary of the mismatch volume are precisely such spectra \cite{OHTA75}. Logvinenko et al. recognised that this is not the case and admitted that imposing such a constraint on the model of the boundary of these objects would result in a loss of precision. Nevertheless, they adopted this model in their algorithm which is summarised in Section \ref{sec:logv_alg}. 

Interestingly, in the earliest work studying the limits of metamerism from the point of view of optimal spectra \cite{ALLEN66, ALLEN69}, Allen stated that optimal spectra on the boundary of the metamer mismatch volume contain "up to 5 regions of 100 percent reflectance, each surrounded by regions of 0 percent reflectance" that is using our wording up to 10 transitions ($m<11$). However, in the next sentence he observed that "in some cases more than five bands may form".

\subsection{Calculating the boundary of the mismatch volume in practice}\label{sec:logv_alg}

Logvinenko et al. proposed the following algorithm for calculation of $\partial\mathcal{M}_5(\mathbf{z}_0,\Phi,\Psi)$ \cite{LOGVINENKO.FUNT.TIP14}. First, a large number of optimal spectra in $\partial\Gamma(\mathcal{O}_5)$ i.e $m<6$ spectra is generated randomly. Then, for each generated spectrum denoted as $r_5(\lambda_1,...,\lambda_5)$ the following optimisation is performed:

\begin{equation}
\min_{\substack{\lambda_1,...,\lambda_5}}||\Phi(r_{5})-\mathbf{z}_0||.
\label{eq:opt5}
\end{equation}

In the next section we will briefly describe our recently published algorithm for calculation of the optimal spectra on the boundary of the object colour solid. What is important about this algorithm is that it does not put any constraints on the number of optimal spectra transitions. This algorithm and most importantly the alternative parametrisation of the optimal spectra it introduces will underpin the main contributions of this paper that is the two algorithms for calculation of the boundary of the mismatch volume not limited by the number of transitions of $\mu$-optimal spectra.

\section{Calculating the boundary of the object colour solid using spherical sampling} \label{sec:calc_OCS}

The key insight in this work \cite{MACKIEWICZ16.PICS} is the observation that the components of vector $\textbf{k}$ in Eq. \ref{eq:ks} have the geometrical meaning i.e. they constitute the normal vector parametrising the boundary of the object colour solid. Since the object colour solid is convex, in the direction $\mathbf{k}$, we can, in closed form, find the unique system response which is maximum. By repeating this procedure for all spherically sampled directions, we can find all points on the OCS.

Formally, we propose the parametric representation of the boundary of the OCS with respect to $\textbf{k}$.

All colour system responses $\Phi(r)=(\phi_1(r),\phi_2(r),...,\phi_N(r))$ from Eq. \ref{eq:col_for2} are projected onto a unit vector $\mathbf{k}$. That is
$$\mathbf{k}\cdot\Phi(r)=\int_{\lambda_{min}}^{\lambda_{max}}r(\lambda)\;\mathbf{k}\cdot\mathbf{s}(\lambda)\,\mathrm{d}\lambda$$. 

It is clear that the maximum value of $\mathbf{k}\cdot\Phi(r)$ is obtained by
\begin{equation}
r_{opt}=r(\lambda;\mathbf{k})=\begin{cases}
0,&\mathbf{k}\cdot\mathbf{s}(\lambda)<0\\
1,&\mathbf{k}\cdot\mathbf{s}(\lambda)\geq 0\\
\end{cases}.
\label{eq:ropt}
\end{equation}

The above observation leads to a very efficient algorithm for calculation of the OCS optimal spectra:

We generate a set of $M$ normal vectors in $R^N$ and store them in the rows of $M\times N$ matrix $\mathbf{P}$ \cite{MARSAGLIA72}. The colour system spectra are stored in $N\times q$ matrix $\textbf{S}$, where  $q$ determines the wavelength resolution e.g. for 1nm resolution, $\lambda_{min}=380$ and $\lambda_{max}=730$, the colour system and reflectance spectra will have 351 components i.e. $q=351$.

We denote a matrix resulting from multiplication of $\mathbf{P}$ by $\mathbf{S}$ as $\mathbf{A}=\mathbf{PS}$. Finally, the signs of the elements of $\mathbf{A}$  determine the set of optimal spectra in matrix $\mathbf{R}$ as:

\begin{equation}
\textbf{R}_{ij} = \begin{cases}
0,&\textbf{A}_{ij}<0\\
1,&\mathbf{A}_{ij}\geq 0\\
\end{cases}
\end{equation}

The final step of the algorithm is to connect the points on the boundary of the OCS into a volume. This can be done using any convex hull algorithm. We have used the Matlab implementation of the Quickhull algorithm \cite{QHULL}.

\begin{figure}[!htb]
\centering
  \includegraphics[width=0.7\columnwidth]{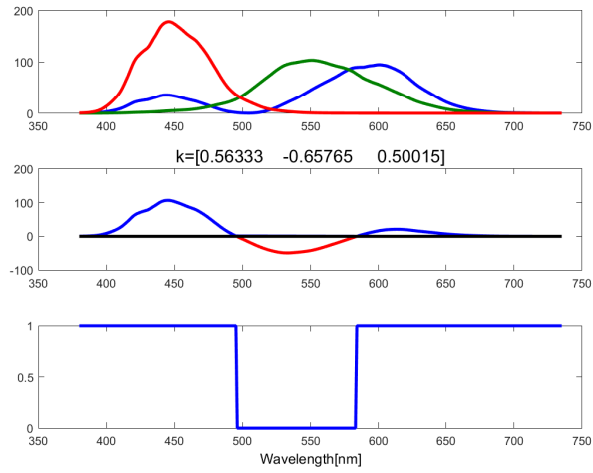}
  \caption{An illustration how to find an optimal spectrum for a given vector $\mathbf{k}$. XYZ colour matching functions multiplied by the D65 illuminant (top). The linear combination (with $\mathbf{k}$ coefficients) of the above three functions (middle). Corresponding optimal spectrum (bottom).}
  \label{fig:findOpt}
\end{figure}

In Fig. \ref{fig:findOpt}, we illustrate with an example how the algorithm works. This example uses the XYZ colour matching functions and the D65 illuminant.

The algorithm is very fast as it does not require optimisation.

We admit that calculation of the object colour solid boundary is not a new problem and there were a number of algorithms proposed in the past. For example, for the set of three sensors, one could generate a number of sensor responses from the set of randomly generated elementary spectra with two transitions \cite{GODAU10, MACKIEWICZ12.CIC}. This said, our method presents two significant benefits. First, the spherical sampling allows for describing the OCS with a small number of samples and second, the normal vector parametrisation allows for optimal spectra with any high number of transitions. Further, this algorithm naturally leads to the related algorithms for calculation of the boundary of the metamer mismatch volume which will be introduced in the next two sections.

\section{Calculating the boundary of the metamer mismatch volume using spherical sampling and linear programming.} \label{sec:calc_lin_prog}

The algorithm presented in the previous section can be used to calculate either $\partial \Phi(\mathcal{X})\in\mathbb{R}^N$  or the larger $\partial \Gamma(\mathcal{X})\in\mathbb{R}^{2N}$. We have stated previously that the metamer mismatch volume $\mathcal{M}$ is a cross-section of $\Gamma(\mathcal{X})$ and it is convex. Then, we can find $\partial\mathcal{M}(\mathbf{z}_0,\Phi,\Psi)$ analogously to the algorithm presented in the previous section. We extremize all spherically sampled directions $\mathbf{k}$ in $\mathbb{R}^{2N}$ subject to $\Phi(r)=\mathbf{z}_0$ and further constraints on the values of the reflectance function. Hence, this optimisation can be written as follows:

\begin{equation}
\max_{\substack{r}} \int_{\lambda_{min}}^{\lambda_{max}}r(\lambda)\mathbf{k}\cdot\mathbf{s}(\lambda),\\
\label{eq:optmy}
\end{equation}
subject to
\[\Phi(r)=\mathbf{z}_0\]
\[0<r(\lambda)<1,\]

where $\mathbf{s}(\lambda)$ are the $2N$ colour system spectra.

We choose a wavelength sampling resolution and write the above optimisation using vector notation as a linear programming problem:

\begin{equation}
\max_{\substack{\mathbf{r}}} \mathbf{(Sk)}^T\mathbf{r},\\
\label{eq:optmy2}
\end{equation}
subject to
\[\mathbf{S}_{\Phi}^T\mathbf{r}=\mathbf{z}_0\]
\[0<\mathbf{r}_i<1 \hspace{1em}\mathrm{for}\hspace{1em} i=1,...,q,\]

where $\mathbf{S}$ is a $q\times 2N$ matrix containing colour system spectra, $\mathbf{S}_{\Phi}$ is a $q\times N$ matrix containing colour system spectra of $\Phi$ and $\mathbf{r}$ is a $q$-vector containing a $\mu$-optimal spectrum in $\partial\mathcal{M}(\mathbf{z}_0,\Phi,\Psi)$ that corresponds to the direction $\mathbf{k}$.

\begin{figure}[!hb]
	\centering
  \includegraphics[width=0.7\columnwidth]{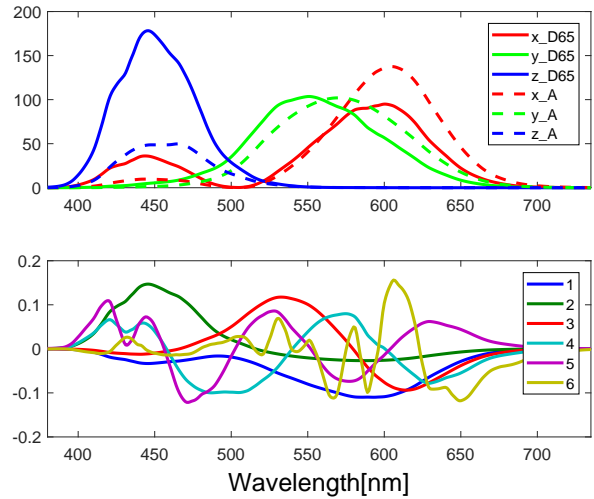}
  \caption{Spectra of colour matching functions under illuminant D65 and A (top) and their orthonormal representation (bottom). The latter are ordered according to their corresponding singular values decrease.}
  \label{Figure:L1L2_Orth}
\end{figure}

We also give an alternative optimisation which as we shall see will offer some improvements over (\ref{eq:optmy2}). This new formulation uses the orthonormal set of colour system spectra which has the potential of achieving a more uniform sampling of the boundaries of both OCS and the metamer mismatch volume. We can write this new optimisation as:

\begin{equation}
\max_{\substack{\mathbf{r}}} \mathbf{(Uk)}^T\mathbf{r},\\
\label{eq:optmy3}
\end{equation}

subject to the same constraints as (\ref{eq:optmy2}), where $\mathbf{U}$ is a $q\times 2N$ matrix containing the set of orthonormal colour system spectra which can be obtained from $\mathbf{S}$ using singular value decomposition \cite{GOLUB12}. In Fig. \ref{Figure:L1L2_Orth} we show six spectra for color matching functions under D65 and A illuminants and their corresponding orthonormal spectra.

Note, that the use of orthonormal spectra ($\mathbf{U}$) with uniform spherical sampling is equivalent to the use of original spectra ($\mathbf{S}$) with certain non-uniform spherical sampling that would emphasise directions in inverse proportion to their corresponding singular values. The singular value decomposition of matrix $\mathbf{S}$ can be written as $\mathbf{S} = \mathbf{UDV}^T$, and thus $\mathbf{U} = \mathbf{SVD}^{-1}$. It is worth noticing that for the above set of six spectra, the ratio of the lowest to the highest singular value is about $10^{-2}$ and hence the resulting spherical sampling can be considered as very much non-uniform.

In Section \ref{sec:exp_res}, we will see that the two versions of the method described in this section do indeed allow for accurate calculation of the metamer mismatch volume theoretical limits. This said, an attentive reader may have spotted that the algorithms presented in this section require a repeated application of linear programming optimisation. If the number of samples is not too high, this may not be a problem. Nevertheless, it is appropriate to ask a question - is there a more efficient algorithm that could avoid optimisation altogether? In the next section, we will show that the answer to this question is `yes' and consequently present an even more efficient alternative algorithm for calculation of the metamer mismatch volumes.

\section{Calculating the boundary of the metamer mismatch volume using spherical sampling without optimisation.}\label{sec:wout_optimisation}

The idea for the calculation of the metamer mismatch volume that would not require optimisation stems from an observation we made in Section \ref{sec:theory}, that is, a 3-D metamer mismatch volume $\mathcal{M}(\mathbf{z}_0,\Phi,\Psi)$ is a particular cross-section of the 6-D object colour solid $\Gamma(\mathcal{X})$. Therefore, we can envisage an algorithm that can be split into two parts - first, building the 6-D OCS and second, calculating its 3-D cross-section. The difficulty here lies in performing these two operations efficiently.

As to the first operation, we have described the algorithm for calculating object colour solids in Section \ref{sec:calc_OCS}. This can be used to obtain their representations, either in 3-D for $\Phi$ or in 6-D for $(\Phi,\Psi)$ responses. Importantly here, this algorithm returns a \textit{vertex} representation of the OCS. While such a representation is perfectly valid, it is not ideal from the point of view of calculating the cross-section of the OCS. Here, we would prefer a \textit{half-space} representation, which as we will later see is particularly suited for our task.

Conversion from vertex to half-space representation is inefficient for a large number of points, particularly in a high dimensional space such as $\mathbb{R}^6$. It requires a convex hull operation to identify all faces followed by the calculation of the surface normals. We expect the number of points required to accurately represent the 6-D OCS to be significantly higher than in the case of the 3-D OCS and therefore conclude that such an approach would be inefficient.

Having said that, there is another way for calculating the half-space representation of the OCS in $\mathbb{R}^6$ that is very efficient. We notice that a vertex $\Gamma$ on the surface of the OCS as calculated using the algorithm in Section \ref{sec:calc_OCS} has a corresponding and known normal vector $\mathbf{k}$ that was used in the calculation of this vertex. Then, the half-space representation of the OCS can be written as $\mathbf{k}_i\cdot \mathbf{x}\leq  b_i$ where $i$ is an index over all spherically sampled normals $\mathbf{k}_i$ and $\mathbf{x}=(\mathbf{z},\mathbf{z}')\in \mathbb{R}^6, \mathbf{z}\in \mathbb{R}^3, \mathbf{z}'\in\mathbb{R}^3 $. The offset of the $i$-th half-space can be calculated using corresponding $\Gamma_i$ as $b_i=\mathbf{k}_i\cdot \Gamma_i$. Finally, the half-space representation of the object colour solid can be written using the following set of inequalities:

\begin{equation}
\mathbf{K}^T\mathbf{x}\leq \mathbf{b}
\label{eq:matrix_half_space}
\end{equation}

where $\mathbf{K}$ is a $6\times t$ matrix containing $t$ spherically sampled unit vectors in its rows and $\mathbf{b}$ is a $t$-vector containing corresponding elements $b_i$.

Notice that this representation overestimates the volume of the OCS as opposed to the earlier vertex representation which underestimated it.

The above has given us the first part of our algorithm that is an efficient to calculate description of the OCS in $\mathbb{R}^6$. In order to calculate the relevant metamer mismatch volume, we need to intersect the 6-D OCS with the affine subspace of dimension 3 defined by $\mathbf{z}=\mathbf{z}_0$ equation. This can be performed efficiently if the OCS is described using the half-space representation.

We know that the solution to the metamer mismatch volume lies in the subspace spanned by the last three coordinate axes and can be written as:

\begin{equation}
\mathbf{x}=\mathbf{Bz'}+\mathbf{x}_0,
\label{eq:solution_subspace}
\end{equation}

where $\mathbf{B}=\left[ \begin{array}{ccc}
0&0&0  \\
0&0&0  \\
0&0&0  \\
1&0&0  \\
0&1&0  \\
0&0&1  \end{array} \right]$ and $\mathbf{x}_0 = \left[ \begin{array}{c}
\mathbf{z}_0\\
0  \\
0  \\
0 \end{array} \right]$.

We substitute (\ref{eq:solution_subspace}) into (\ref{eq:matrix_half_space}) and obtain the half-space representation of the metamer mismatch volume:

\begin{equation}
\mathbf{K}^T\mathbf{Bz'}\leq \mathbf{b}-\mathbf{K}^T\mathbf{x}_0.
\label{eq:mmv_half_space}
\end{equation}

In order to calculate the vertices (and the volume) of the metamer mismatch volume, we need to perform the half-space intersection operation. This can be done using tools such as Qhull \cite{QHULL} and the duality transform of Preparata and Shamos\cite{PREPARATA}. We expect this operation to be efficient as it is performed in a low dimensional space ($\mathbb{R}^3$) and also we predict that only a limited number of planes in (\ref{eq:mmv_half_space}) will bind the intersection. 

\section{Experiments and Results} \label{sec:exp_res}

We performed a number of experiments testing the two algorithms presented in the previous sections. First, we tested the two variants of the algorithm that uses an optimisation and compared them with the algorithm presented in \cite{LOGVINENKO.FUNT.TIP14} that constitutes our benchmark. In all experiments we used CIE 1931 2-deg, XYZ colour matching functions modified by Judd and Vos \cite{VOS78}. We have built metamer mismatch volumes for the change of illuminant for three illuminants: D65, A and F11 \cite{HUNT11} for all 6 pairs. Below, we will only show the results of the two conditions for the change of illuminant from D65 to A and F11 to D65, which are representative of all the results.

\begin{figure}[!hb]
	\centering
  \includegraphics[width=0.7\columnwidth]{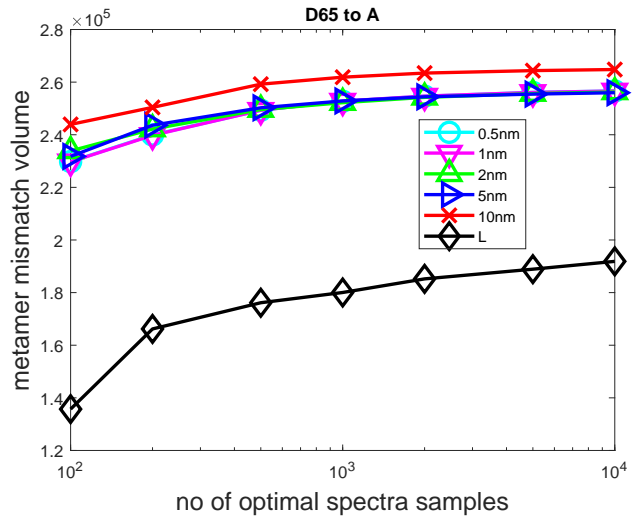}
  \caption{Metamer mismatch volumes calculated using the method presented in Section \ref{sec:calc_lin_prog} for the flat grey reflectances with 50\% reflectivity for the two sets of orthonormal colour system spectra for the change of illuminant from D65 to A with spectral sampling varying from 0.5nm to 10nm. (L) - method proposed by Logvinenko et al. \cite{LOGVINENKO.FUNT.TIP14}. Results for 0.5nm and 1nm sampling resolutions are almost the same and hence 0.5nm is hidden under the 1nm plot.}
  \label{Figure:D652A_1}
\end{figure}

\begin{figure}[!hb]
	\centering
  \includegraphics[width=0.7\columnwidth]{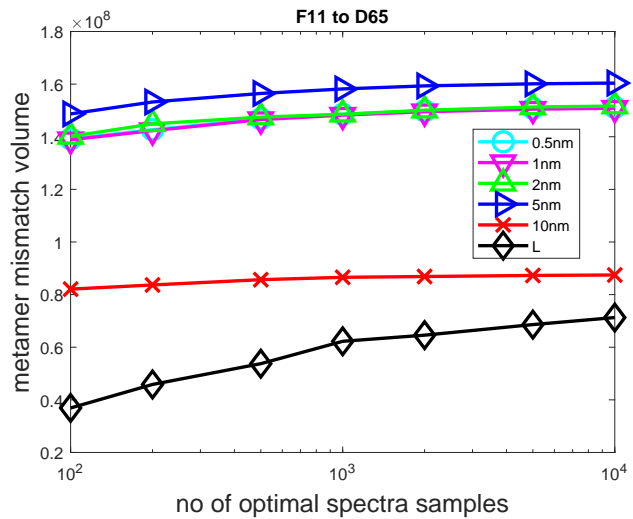}
  \caption{As in Fig. \ref{Figure:D652A_1}, but for the change of illuminant from F11 to D65.}
  \label{Figure:F112D65_1}
\end{figure}

In our first experiment, we wished to determine the appropriate wavelength sampling for our algorithm. We tested wavelength resolutions from 0.5nm to 10nm. The metamer mismatch volumes were calculated for the flat grey reflectance $(r_0=0.5)$ for the increasing number of samples generated using spherical sampling as described in Section \ref{sec:calc_lin_prog}. Here, we used the variant of the algorithm employing linear programming and the orthonormal colour system spectra. We also compared the volumes generated with this algorithm with those obtained using \cite{LOGVINENKO.FUNT.TIP14}. Note that the Logvinenko et al.'s method uses parametrisation that does not require us to choose the wavelength sampling resolution. The results of this experiment can be seen in Figures \ref{Figure:D652A_1} and \ref{Figure:F112D65_1}.

In Fig. \ref{Figure:D652A_1}, we can see that the volumes obtained for wavelength sampling resolutions of 0.5nm and 1nm are almost identical. The 2nm and 5nm are very similar, whereas the 10nm sampling significantly overestimates the volume. These observations are confirmed in Fig. \ref{Figure:F112D65_1}, where again we can see that 0.5 and 1nm sampling produce almost identical results. However, this time the errors for wavelength resolutions above 1nm become more visible as this result has been produced for the change of illuminant involving F11, which has a spectrum that is less smooth than both D65 and A and hence may require a lower sampling resolution. Therefore, we will use the wavelength resolution of 1nm in all subsequent experiments.

Figures \ref{Figure:D652A_1} and \ref{Figure:F112D65_1} also show that the metamer mismatch volumes obtained for the Logvinenko et al. algorithm are significantly smaller (from approximately 25\% to 70\%). This is particularly visible for a small number of samples and when the F11 illuminant is involved.

\begin{figure}[!htb]
	\centering
  \includegraphics[width=0.7\columnwidth]{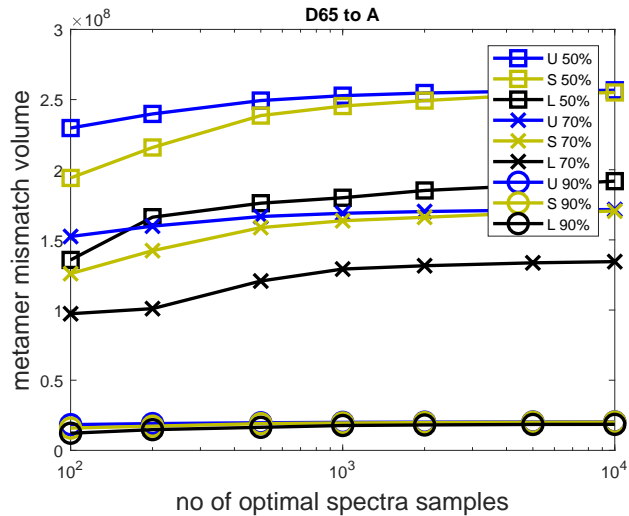}

  \caption{Metamer mismatch volumes calculated using the method presented in Section \ref{sec:calc_lin_prog} for the flat grey reflectors with 50\%, 70\% and 90\% reflectance for the the change of illuminant from D65 to A using the orthormal colour system spectra (U), original colour system spectra (S) with 1nm spectral sampling. (L) - method proposed by Logvinenko et al. \cite{LOGVINENKO.FUNT.TIP14}.}
  \label{Figure:D652A_2}
\end{figure}

\begin{figure}[!htb]
	\centering
  \includegraphics[width=0.7\columnwidth]{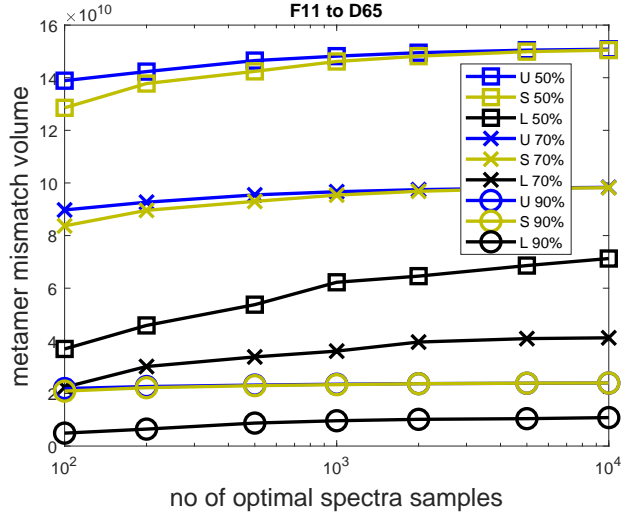}
  \caption{As in Fig. \ref{Figure:D652A_2}, but for the change of illuminant from F11 to D65.}
  \label{Figure:F112D65_2}
\end{figure}

Figures \ref{Figure:D652A_2} and \ref{Figure:F112D65_2} show the results of the second experiment where we investigated two further aspects of the linear programming algorithm. First, we analyse the volumes produced by the two versions of the algorithm, for the original and orthonormal colour system spectra. Second, we look at the size of the mismatch volumes along the achromatic line for both variants of the algorithm. As to the first aspect of this experiment, we can see that the orthonormal colour system spectra indeed result in better distribution of the samples describing the mismatch volume and consequently produce larger volumes. More specifically, in Fig. \ref{Figure:F112D65_2} we can see that for the flat grey reflectance (50\%), the version of the algorithm utilising the orthonormal spectra with 100 spherical samples produces the mismatch volume estimate that is matched only by the original version of the algorithm with 500 samples.

Regarding the second aspect of the experiment, we can see that as expected the metamer mismatch volumes are the largest in the centre of the OCS. Moreover, we can also see that the Logvinenko et al.'s method significantly underestimates the sizes of the mismatch volumes particularly in the centre of the OCS.

The prior art method \cite{LOGVINENKO.FUNT.TIP14} uses an inefficient strategy of random initialisation of the transition wavelengths which then tend to converge to certain clusters on the surface of the mismatch volume. Hence, this method requires a very large number of 5-transition optimal spectra - preferably above 10000 samples - in order to estimate the size of $\mathcal{M}_5$, which we have reasons to expect will still significantly underestimate the size of $\mathcal{M}$. On the other hand, the better performing version of our algorithm usually requires as few as 1000 samples to accurately estimate the volume of the larger mismatch volume $\mathcal{M}$.

\begin{figure}[!hb]
	\centering
  \includegraphics[width=0.7\columnwidth]{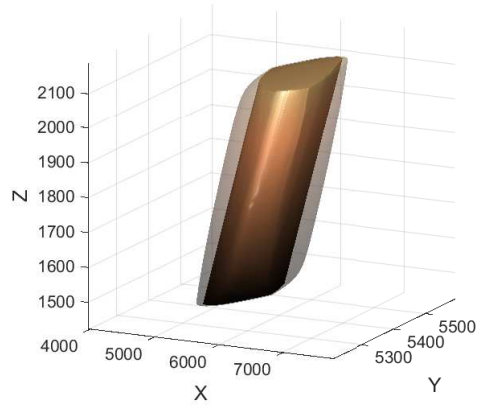}
  \caption{Comparison of the metamer mismatch volumes calculated using the method presented in Section \ref{sec:calc_lin_prog} for the flat grey reflectance with 50\% reflectance for the change of illuminant from D65 to A using 1nm spectral sampling with the corresponding volume calculated by the method proposed in\cite{LOGVINENKO.FUNT.TIP14}. Both methods use 10000 samples. Plotted in the CIE XYZ colour space.}
  \label{Figure:D652A_3}
\end{figure}

\begin{figure}[!htb]
	\centering
  \includegraphics[width=0.7\columnwidth]{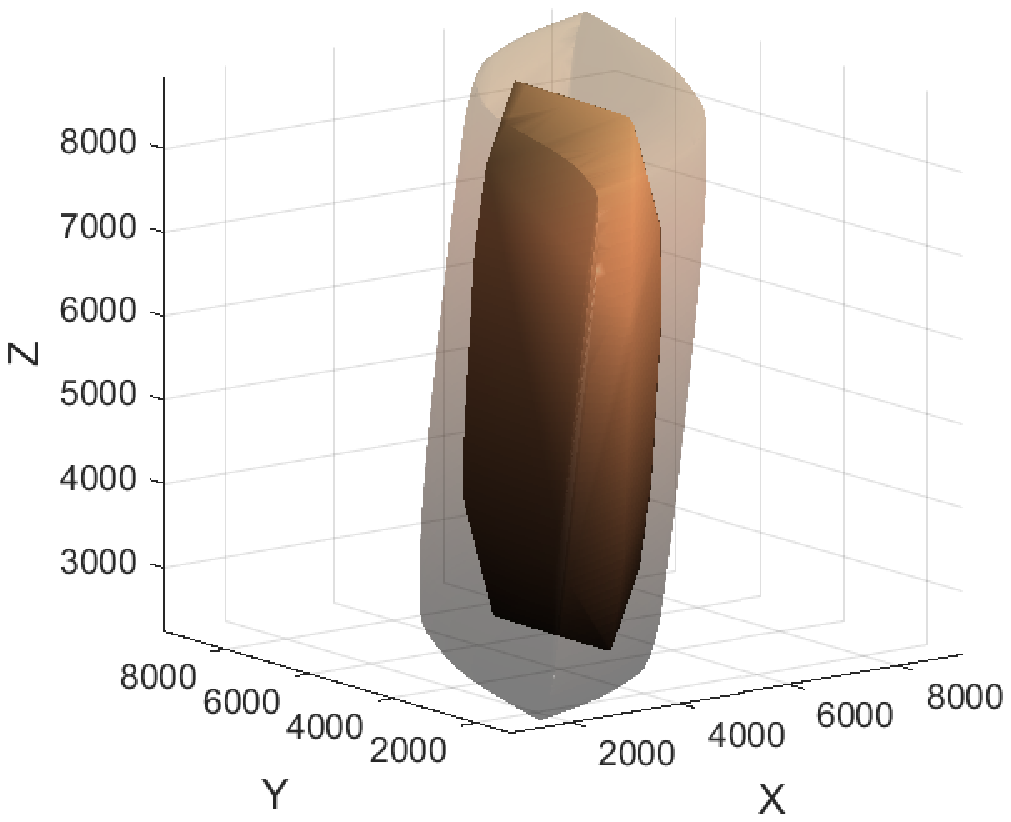}
  \caption{As in Fig. \ref{Figure:D652A_3}, but for the change of illuminant from F11 to D65.}
  \label{Figure:F112D65_3}
\end{figure}

\begin{figure}[!htb]
	\centering
  \includegraphics[width=0.7\columnwidth]{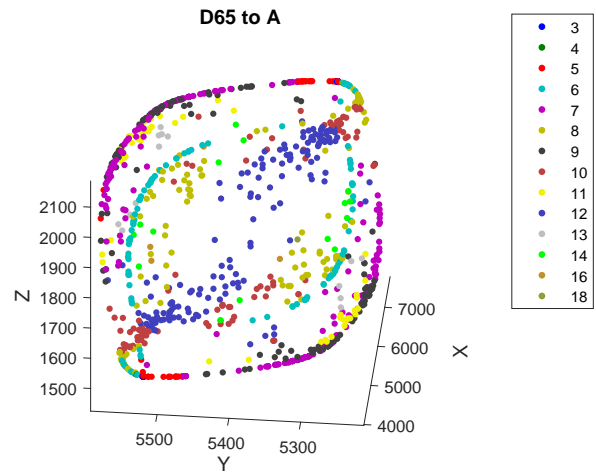}
  \caption{Random subset of 1000 points on the surface of the larger convex hull in Fig. \ref{Figure:D652A_3}. The points are coloured according to the number of transitions of the corresponding optimal spectra.}
  \label{Figure:D652A_4}
\end{figure}

\begin{figure}[!htb]
	\centering
  \includegraphics[width=0.7\columnwidth]{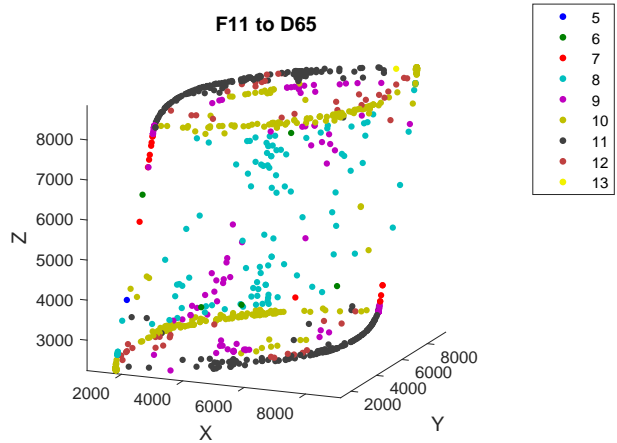}
  \caption{As in in Fig. \ref{Figure:D652A_4}, but for the larger convex hull in Fig. \ref{Figure:F112D65_3}.}
  \label{Figure:F112D65_4}
\end{figure}

In Figures \ref{Figure:D652A_3} and \ref{Figure:F112D65_3}, we can see the graphical comparison of the mismatch volumes produced by our algorithm and the Logvinenko et al. method. The mismatch volume approximation $\mathcal{M}_5$ is indeed clearly contained within the theoretical limits of the mismatch volume $\mathcal{M}$ produced by our algorithm. The reasons behind this phenomena can be better understood from Figures \ref{Figure:D652A_4} and \ref{Figure:F112D65_4}. There, we show the same convex hulls that were created using our algorithm, but this time we highlight with colour the number of transitions of the corresponding optimal spectra. Both figures use 1000 random samples on the surface of each volume. We can see that for the change of illuminant from D65 to A the number of transitions on the surface of the metamer mismatch volume ranges from 3 to 18 and for the F11 to D65 from 5 to 13. We can also see that the optimal spectra with the same number of transitions form regions or curves on the surface of these objects.

\begin{figure}[!htb]
	\centering
  \includegraphics[width=0.7\columnwidth]{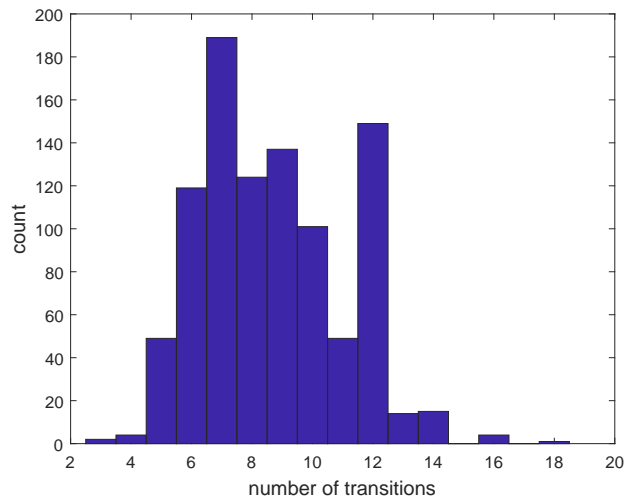}
  \caption{A histogram illustrating the frequency of occurrence for different number of transitions in optimal spectra. Corresponds to data in Fig. \ref{Figure:D652A_4}.}
  \label{Figure:D652A_5}
\end{figure}

\begin{figure}[!htb]
	\centering
  \includegraphics[width=0.7\columnwidth]{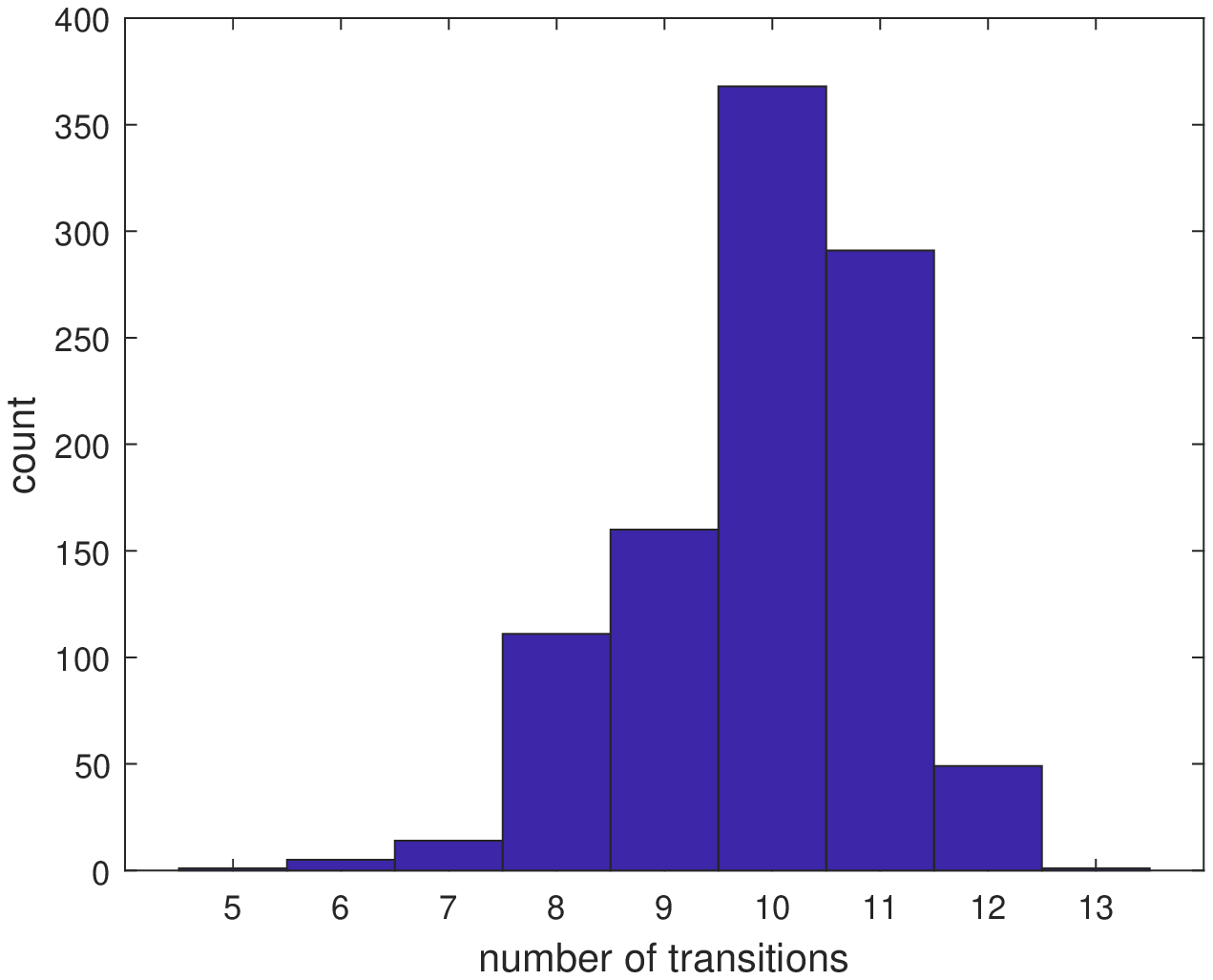}
  \caption{As in in Fig. \ref{Figure:D652A_5}, but for the change of illuminant from F11 to D65. Corresponds to data in Fig. \ref{Figure:F112D65_4}}
  \label{Figure:F112D65_5}
\end{figure}

Histograms showing the frequency of occurrence of optimal spectra with different number of transitions can be seen in Figures \ref{Figure:D652A_5} and \ref{Figure:F112D65_5}. They clearly show that the vast majority of optimal spectra have a higher number of transitions than 5.  The low number of optimal spectra of less than or equal five transitions for the change of illuminant from F11 to D65 is particularly striking which explains why for this condition there is such a large difference between the volumes produced by our algorithm and the prior art.

\begin{figure}[!htb]
	\centering
  \includegraphics[width=0.7\columnwidth]{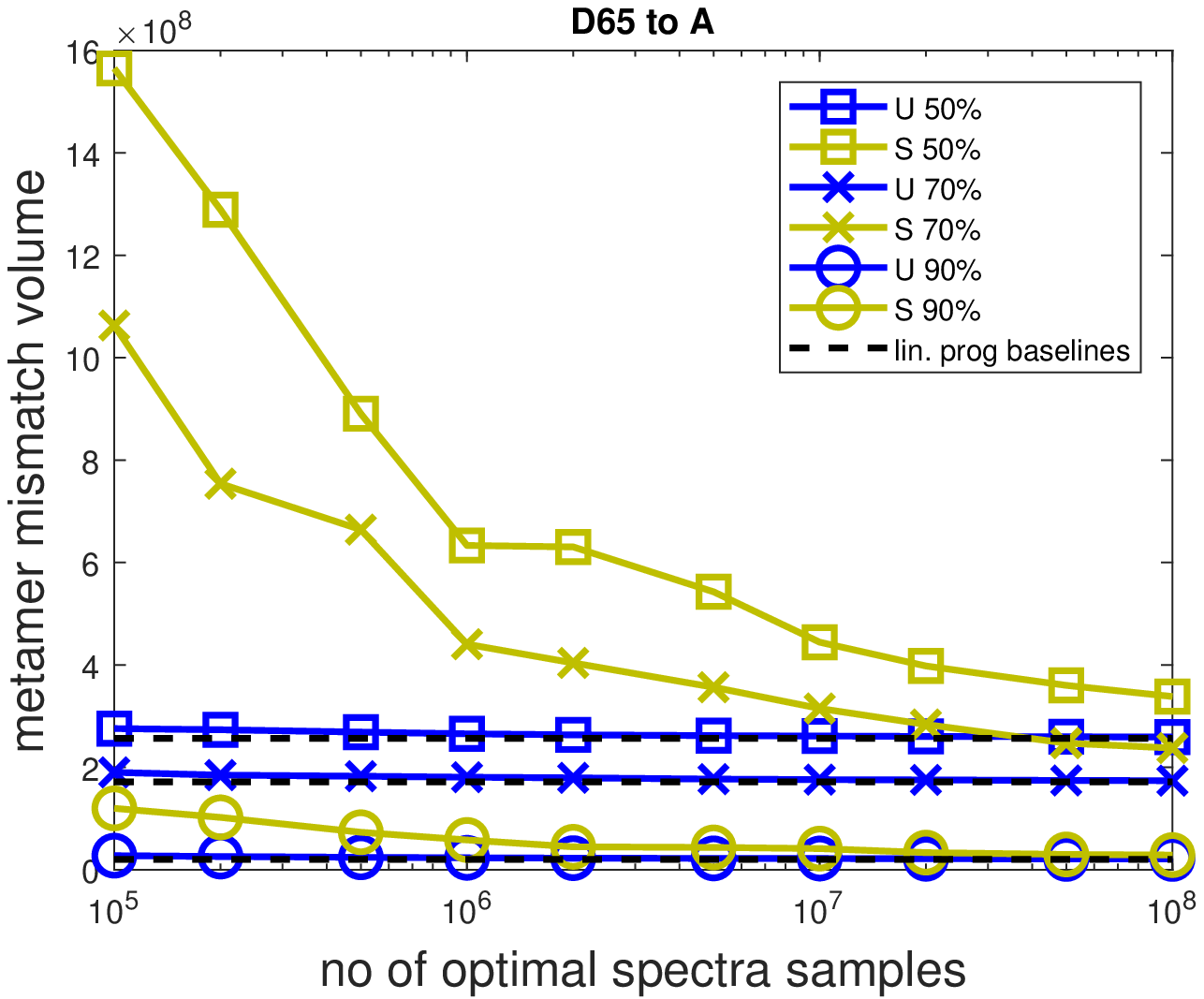}
  \caption{Metamer mismatch volumes calculated for the flat grey reflectors with 50\%, 70\% and 90\% reflectance using the method presented in Section \ref{sec:wout_optimisation} for the the change of illuminant from D65 to A using the orthonormal colour system spectra (U), original colour system spectra (S) with 1nm spectral sampling. Dashed lines correspond to the respective volumes (for $10^4$ samples) in Fig. \ref{Figure:D652A_2}.}
  \label{Figure:D652A_6}
\end{figure}

\begin{figure}[!htb]
	\centering
  \includegraphics[width=0.7\columnwidth]{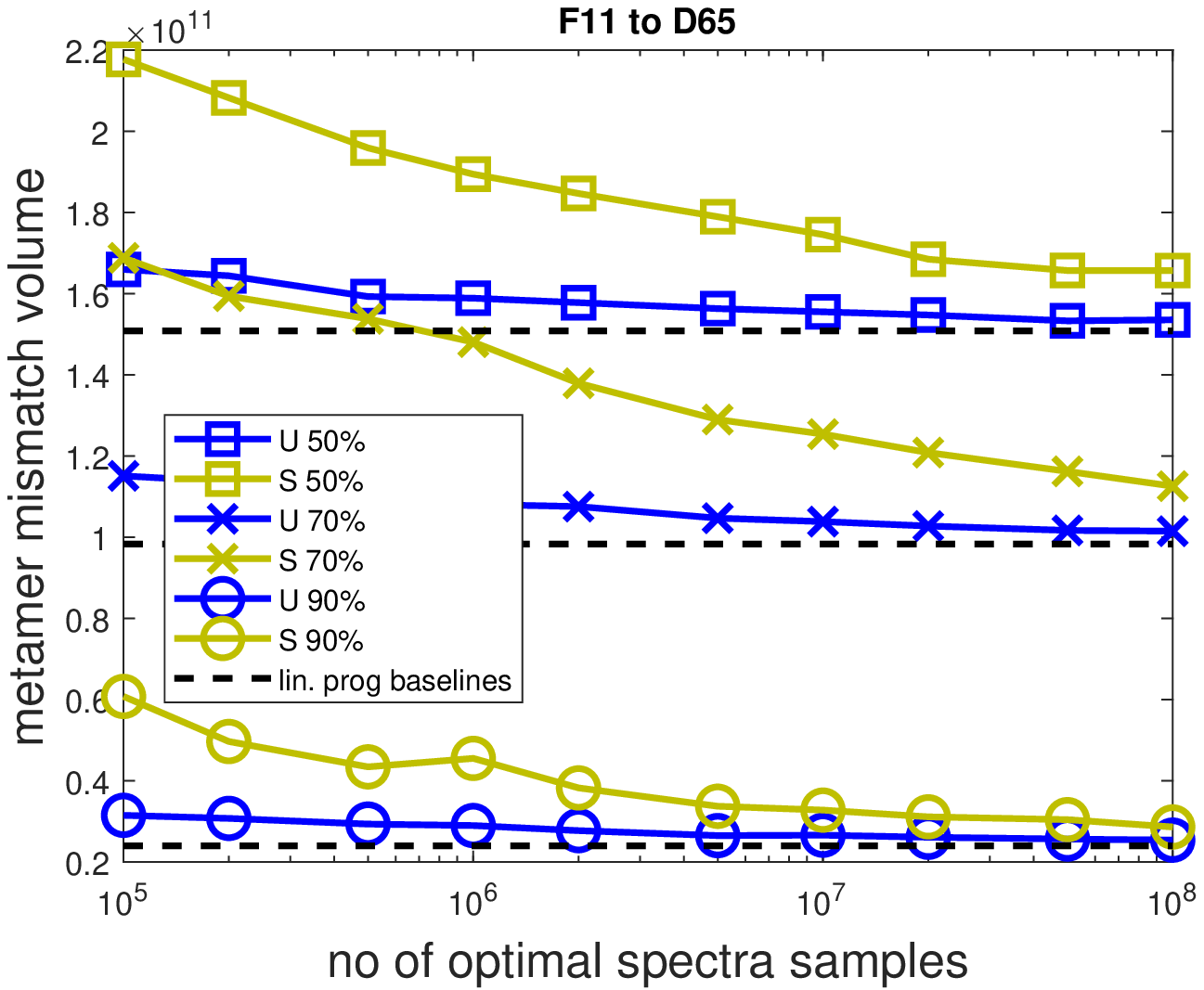}
  \caption{As in Fig. \ref{Figure:D652A_6}, but for the change of illuminant from F11 to D65. Dashed lines correspond to the respective volumes (for $10^4$ samples) in Fig. \ref{Figure:F112D65_2}.}
  \label{Figure:F112D65_6}
\end{figure}

In the final experiment, we calculated metamer mismatch volumes using our final method requiring no optimisation that was presented in Section \ref{sec:wout_optimisation}. Here, we also used 1nm sampling. We present our results for the same two conditions: change of illuminat from D65 to A (see Figure \ref{Figure:D652A_6}) and from F11 to D65 (Figure \ref{Figure:F112D65_6}). Analogously to the previous experiment, we plot the volumes for the flat grey reflectance of 50\%, 70\% and 90\% reflectance and using two sensor sets: original and orthonormal. We compare the metamer mismatch volumes to those obtained in the previous experiment for the largest number of samples ($10^4$, see Figs \ref{Figure:D652A_2} and \ref{Figure:F112D65_2}) - these are plotted as black dashed lines. 

A reader should notice that the range of spherically sampled points we tested is substantially different comparing to the previous experiment. Earlier, we stopped at $10^4$ samples and here we performed our calculations for the range of samples from $10^5$ to $10^9$. In Section \ref{sec:wout_optimisation}, we predicted that the number of samples needed to cover the 6-D OCS will be significantly higher than for the 3-D mismatch volume and this is reflected in our figures.

Our next observation from these figures is that as predicted the approximated mismatch volumes are overestimated and as the number of samples increases, the volumes tend to those obtained using our earlier linear programming method.

An interesting observation that is different to what we have seen earlier is the difference in performance for the two sensor sets. While we have seen better approximation with orthonormal sensors for the linear programming method, here the gap in performance of the two sensor sets is much wider, particularly for the change of illuminant from D65 to A where using the orthonormal sensors the volumes are more accurately represented with as few as $10^5$ samples than using the original sensor set with as many as $10^8$ samples.

As to the change of illuminant from F11 to A (see Fig. \ref{Figure:D652A_2}), the orthonormal sensors also provide much better approximations of real mismatch volumes. This said, here we need to use a larger number of samples to obtain a comparable accuracy.

Our final comment concerns the major advantage of the method discussed here, namely speed of execution. Linear programming approach required approximately 500s to calculate the metamer mismatch volume using $10^4$ samples and proportionally less time if less points were required. The method presented here is much faster. First, we note that the algorithm can be split into two parts: the calculation of the 6-D object colour solid for a given colour system and the calculation of the intersection with the 3-D affine subspace corresponding to a required metamer. The latter takes less time than the former. If an application requires a number of mismatch volumes to be calculated for the same colour system, then we need to calculate the 6-D OCS only once. This step takes approximately 170s for $10^8$ samples and proportionally less if less sampling points are required i.e. approximately 1.7s for $10^6$ samples. The second step is faster than the first and takes approximately 70s and 0.7s for $10^8$ and $10^6$ samples respectively. 

While different conditions may require different numbers of samples for any of the two methods, it is clear that the half-space intersection method is significantly faster e.g. for the change of illuminant from D65 to A, an accurate mismatch volumes can be calculated in less than a second for $10^6$ samples, whereas an approximation of this volume with $10^3$ samples using linear programming would require 50s. All experiments were performed on a PC running Matlab 2014a on Intel i7-4790 at 3.6GHz and 32GB of memory.

\section{Conclusions}

Metamer mismatching is an important phenomenon in colour science. Here, we proposed two novel algorithms for calculation of the theoretically maximal metamer mismatch volumes. To our knowledge, they are the first algorithms capable of calculating a precise maximum extent of these volumes. Our figures when compared with those produced by the earlier methods show that the 5-transition approximation results in significantly smaller mismatch volumes (sometimes above 50\%!). Our two algorithms are computationally efficient
due to simple formulations (either as a linear programming optimisation or as a half-space intersection) and a relatively small number of spherical samples required to provide precise mismatch volume estimates. Importantly, we note that both algorithms take advantage of the small number of samples as a result of their use of orthonormal sensors which is equivalent to a certain non-uniform sampling of these objects.

\end{document}